\documentstyle[psfig]{l-aa}

\begin{document}

\newcommand{\arcm}{\hbox{$^\prime$}}
\newcommand\uarcs{\hskip-0.27em\arcsec\hskip-0.02em}

\thesaurus{03        
          (03.01.2;  
           03.13.2;  
           03.13.5;  
           11.06.2;  
           11.16.1;  
           )}        %

\title{The $t$ system: 
a new system for estimating the total magnitudes of galaxies}


\author{Christopher Ke-shih Young\inst{1},
        Nigel Metcalfe\inst{2},
        Jin Zhu\inst{1},
        Hong Wu\inst{1} and 
        Jian-sheng Chen\inst{1} }

\offprints{C.K.\ Young by e-mailing c.young1@physics.oxford.ac.uk}

\institute{
$^{1}$Beijing Astronomical Observatory, Chinese Academy of Sciences, Beijing 100080, China\\
$^{2}$Department of Physics, University of Durham, South Road, Durham DH1 3LE, England
          }

\date{Accepted 1997 November 12}

\maketitle

\markboth{C.K.\ Young et al.: $t$-system total magnitudes}{}

\begin{abstract} 
We present a new, but simple, procedure for estimating the total magnitudes of 
galaxies. This procedure involves the out-focusing of digital galaxy images 
numerically, the fitting of the resulting surface-brightness profiles with a 
{\em single} generalised profile model and the extrapolation of the fitted 
profiles to infinite radial distances. 
This new system, which we denote $t$, differs fundamentally from the $T$ 
system (of the {\em Reference Catalog of Bright Galaxies\/} series) in that: 
(1) it enables a galaxy's luminosity profile to be extrapolated without the 
need for any prior morphological classification, and (2) it is applicable to 
images of widely different spatial resolutions (including unresolved ones) 
because it takes into account systematic effects due to differential image 
resolution.
It also differs fundamentally from the Kron system in that: (1) it can be
derived directly from surface photometry without the need to go back to the 
plate scans or CCD frames (unless the surface photometry is of high
resolution and/or the galaxies being measured are very bright), and (2) it 
can cope with merged images (provided they are separable by image-segmentation 
software).
Through worked examples, we demonstrate the stability of $t$-system total 
magnitudes with respect to morphological type, the seeing conditions at the 
time of observation, degree of smoothing and limiting isophote. We also 
compare and contrast the new system with both the $T$ system and the Kron 
system, and investigate the advantages and limitations of each of the three 
systems.
   \end{abstract}

\section{Introduction}

At sufficiently large angular distances from the centre of any galaxy image, 
the surface-brightness contribution due to the galaxy becomes, at some point, 
indistinguishable from the surface-brightness of the surrounding sky. 
The limit on reliable observation 
is determined primarily by the noise and often corresponds to those points at 
which the galaxy's surface brightness has fallen to several percent of the sky, 
though limits in the region of 0.1\% of the sky are measurable on occasions. 

Although total magnitudes are required for many astrophysical applications, it
is therefore not possible to measure them directly. Instead, estimates are 
normally obtained by means of extrapolating model profiles fitted to those 
parts of the galaxy-light profiles (whether surface-brightness or integrated) 
that can be measured reliably. However, galaxies of different morphological 
type have very different profile shapes, and a wide range of models have 
generally had to be invoked. In Kron's (1980) system though, the flux due
to a galaxy is measured to {\em very\/} large radial distances, so that 
different models do not need to be invoked. Such a procedure has its advantages,
but at the very large radial distances involved, the signal due to the galaxy
is often such a small fraction of the noise that large random errors cannot
be avoided.

An alternative approach to profile extrapolation is $very$-low-resolution 
imaging so that, in theory at least, all target galaxies essentially become 
point sources and yield images of almost identical structure. Total 
magnitudes then become a simple function of the full-width half maximum (FWHM) 
of the image point-spread function. In practice, the FWHM of the point-spread 
function has to be much larger than the intrinsic angular sizes of the 
target galaxies for this method to yield reliable magnitudes. However, the
wider the point-spread function is, the lower the mean surface-brightness of
each image becomes and the greater the errors due to the noise become. 
In the compilation of their {\em Catalog of Galaxies and Clusters of Galaxies\/}, 
Zwicky et al.\ (1961, 1963, 1965, 1966, 1968) measured their total magnitudes 
visually from {\em out-of-focus\/} photographic plates. Out-of-focus images
generally have a complicated point-spread function, but as long as they
exhibit little variation across a single plate or CCD frame, this should not be a
problem.

The new system presented in this paper combines both approaches. We believe
that it can yield very large numbers of reliable total-magnitude measurements
very efficiently, and that it is therefore particularly suitable for 
galaxy-survey work.

\section{Existing extrapolation methods}

In his pioneering study, Hubble (1930) found that the surface-brightness 
profiles of elliptical galaxies (which he measured along either the major or 
minor axes) seemed to be well fitted by the law:

\begin{equation}
\sigma(r) = \frac{\sigma_{H}}{(1 + \frac{r}{r_H})^{2}},
\end{equation}

\noindent
where an elliptical-galaxy profile can be uniquely described by two terms:
$\sigma_{H}$ (a central surface brightness) and $r_{H}$ (a scale length).
The main problem with this law is that its integrand with respect to $r$
diverges as $r$ increases, and it is therefore unsuitable for extrapolation
to large $r$. 

An alternative representation was proposed by de Vaucouleurs (1948):

\begin{equation}
\sigma(r) = \sigma_{e} \; \mbox{dex} \;
\{-3.33[(\frac{r}{r_{e}})^{\frac{1}{4}}-1]\},
\end{equation}

\noindent
where an elliptical-galaxy profile can be uniquely specified by  $r_{e}$
(the effective radius which contains half the galaxy's light). $\sigma_{e}$
is the surface brightness at $r=r_{e}$. This representation has the advantage 
of a convergent integrand: 

\begin{equation}
2 \pi \int_{o}^{\infty} \sigma(r) r \mbox{ } dr = 22.4 \sigma_{e} {r_{e}}^{2},
\end{equation}

\noindent
However, it has the disadvantage that strictly, the total luminosity needs to be
known before the effective parameters can be evaluated. 

A move away from wholly empirical functions was made by King (1966). King's
models were constructed with tidally truncated globular star clusters in mind
but have also been applied to elliptical galaxies. They assume that at all
points within a stellar system, the frequency distribution of stars as a
function of position and velocity can be described by an isothermal Gaussian
minus an offset. At any position within the system, the offset is chosen so as
to ensure that the frequency distribution is zero for stars with velocities
equal to and in excess of the local escape velocity. This succeeds in reducing
an otherwise infinite isothermal space distribution of stars to one with a
finite radius. Nevertheless, the empirical law of de Vaucouleurs actually 
offers a better description of tidally unperturbed galaxies than does King's 
model.

Despite the intricate structure exhibited by many spiral galaxies, it has long
been realised, e.g. by de Vaucouleurs (1958) that their smoothed light profiles 
could be separated into two components: a central component approximately obeying an
$r^{\frac{1}{4}}$ law\footnote{
de Jong (1996) has recently shown that bulge surface-brightness profiles actually obey 
a range of laws in which the index to which $r$ is raised varies considerably and
is often much higher than 0.25.}
(corresponding to the spheroidal bulge, denoted $s$) 
and a more extended component approximately obeying an exponential law 
(corresponding to the disc, denoted $d$):

\begin{eqnarray}
\lefteqn{\sigma(r)=} \nonumber \\
 & & \sigma_{e,s} \mbox{dex}
\{-3.33[(\frac{r}{r_{e,s}})^{\frac{1}{4}}-1]\} \; + \;
\sigma_{0,d} \exp [-(\frac{r}{r_{0,d}})],
\end{eqnarray}

\noindent
where $\sigma_{0,d}$ (the central surface-brightness due to the disc alone)
and $r_{0,d}$ (the scale length of the disc component) uniquely describe the
disc component. The contribution due to this exponential component decreases
as one progresses from late-type spirals to earlier types, but is not completely
absent from lenticular objects or even classical ellipticals.

This trend in galaxy-profile characteristics from early through late 
types was exploited by de Vaucouleurs et al.\ (1976, 1991) 
during the compilation of their {\em Second\/} and {\em Third 
Reference Catalogs of Bright Galaxies\/} (hereunder RC2 and RC3 respectively).  
Their elaborate and widely-used scheme for extrapolating both aperture and 
surface-photometry measurements of galaxies, 
in order to estimate total magnitudes, is known as the $T$ system. 
Although the $T$ system has been used successfully to extrapolate the
profiles of well resolved images of classical galaxies, 
its applicability to dwarf galaxies (as well as 
ellipticals intermediate between true dwarfs and true classicals) 
and to low-resolution images (of all galaxy types) is questionable.
There appear to be at least three major limitations to the $T$ system, 
the main consequences of which are summarised by Young (1997) and 
will be dealt with in more detail by Young et al.\ (in preparation).\\
(1) The scheme does not take into account the possibility that galaxies with
profiles steeper than exponentials exist.
As is evident from Young \& Currie (1994, 1995) many dwarfs have profiles
that {\em are\/} steeper than exponentials and some even have profiles as steep
as Gaussians. These objects are therefore beyond the scope of the $T$ system
which necessarily over-estimates their luminosities.\\
(2) A morphological classification must be attempted first in order to be able 
to select the most appropriate extrapolation model for each galaxy concerned. 
This is normally done by eye. Many non-classical elliptical galaxies 
(including some with some characteristics of irregulars) e.g.\ IC~3475, 3349, 
3457 and 3461, were classified as $r^{\frac{1}{4}}$-law objects by de 
Vaucouleurs et al.\ (1976, 1991) and their profiles extrapolated accordingly. 
In fact IC~3475 has a very exponential profile as demonstrated by Vigroux et 
al.\ (1976), and the other three objects listed have profiles slightly steeper 
than an exponential as demonstrated by Young \& Currie (in press). This means that
whilst these objects could be accommodated by the $T$ system if it were treated
as objects with exponential profiles, their luminosities were in fact
severely over-estimated by of the order of 100\% in both the RC2 and the RC3 
(Young, 1997; Young et al., in preparation) because of limitations in the 
morphological typing procedure. The need to estimate the profile shape by eye 
before an extrapolation can be performed is therefore a serious short-coming 
of the $T$ system.\\
(3) No account is taken of atmospheric or instrumental effects that degrade
the resolution of a galaxy image and thereby modify its {\em measurable\/}
surface-brightness profile. Clearly a low-resolution image of a particular
galaxy is likely to have a more Gaussian-like profile than is a high-resolution
image of the same galaxy. This makes the $T$ system difficult to apply
consistently to images of different resolution and even to images of almost
identical galaxies at widely differing distances.

A worthy alternative to the $T$ system has long been the Kron (1980) system,
in which the rate of growth of the signal with respect to the signal itself is 
considered as a function of radial distance from the centre of each galaxy 
image. The light is then measured within a circular region of a radius
corresponding to the point at which the logarithmic derivative of the light
growth curve becomes smaller than the {\em same\/} upper limit for all 
target galaxies. In practice there are strong constraints on the suitable 
values for this limit, and the range of suitable values generally require an 
aperture of radius equal to about twice the effective (half-light) radius. 
The fraction of the total light within this aperture (which in practice is 
typically of the order of 95\%) is then assumed to be constant for all 
galaxies, and total magnitudes can be obtained by extrapolating the 
measured aperture magnitude values by the same amount (which in practice 
simply means adding typically about $-$0.05 mag.\ to each aperture magnitude).

The Kron system has two major advantages over the $T$ system, namely that the
magnitude measurement process is independent of galaxy morphological type and 
that atmospheric effects that degrade image resolution and distort galaxy 
luminosity profile shapes are taken into account. However, it does have
at least three major drawbacks.\\
(1) Large random errors are present due to having to measure luminosity 
growth curves and [even harder] their derivatives, out to very large radial 
distances where the signal due to the galaxy is a very small fraction of the 
noise. Note that the Kron system does not take the signal-to-noise ratio as
a function of radial distance into account at all. This can be a particularly 
serious problem when dealing with low-surface-brightness galaxies such as 
dwarfs.\\
(2) The need for very large apertures restricts which objects can be measured
in crowded fields when undesirable galaxy or stellar images lie adjacent to
the target objects. For this reason the Kron system has mainly been applied
to field galaxies rather than cluster ones.\\
(3) Unlike the $T$ system which is based on aperture and surface photometry
measurements that can be extracted from the literature in their published
form, one cannot compute Kron-system total magnitudes without access to the
original digital-image data.

Despite the existence of the elaborate $T$ and Kron systems,
most large machine surveys of galaxies have, understandably, adopted 
simpler algorithms for extrapolating large numbers of isophotal magnitudes to 
totals. In the APM Southern-sky Survey of Maddox et al.\ (1990) the majority 
of galaxy images were assumed to be seeing dominated, and therefore to have 
approximately Gaussian profiles (on average at least). This was a particularly 
efficient method, as knowledge of an isophotal magnitude and the angular area 
of that isophote was sufficient to specify the parameters of a Gaussian profile 
uniquely; as described in detail in Maddox et al.\ (1990). However, such an 
approach is only applicable to unresolved galaxy images from sufficiently 
small and/or sufficiently distant objects. 

In addition to the extrapolation systems discussed in this section, there
are of course others. However, such systems have generally only been applied to 
limited galaxy samples of a particular morphological type.

\section{The $\lowercase{t}$ system}

This system was first adopted by Young (1994) and Young \& Currie (in press) in the compilation of their 
{\em Virgo Photometry Catalogue\/} (hereunder VPC); as well as by 
Young \& Currie (1995), Drinkwater et al.\ (1996) and Young (1997);
who all quote $t$-system total magnitude values from the VPC.
The $t$ system is a system for extrapolating surface-brightness profiles of 
sufficiently low resolution, to $r=\infty$ where $r$ is reduced radial 
distance ($\sqrt{r_{major}r_{minor}}$). This means that high-resolution 
images (those of non-nucleated dwarf and intermediate ellipticals excepted) must be smoothed 
sufficiently {\em prior to\/} any surface-brightness profile being 
parametrized, but has the important advantage that low-resolution images
can be measured even if their profiles are significantly distorted by e.g.\ 
seeing effects or poor sampling. Although alternative smoothing functions
can be used, we recommend a [radially-symmetric two-dimensional] Gaussian
function.

In order to generate an extrapolated total $t$-system magnitude from a low-resolution 
surface-brightness profile, S\'{e}rsic's (1968) law is adopted: 

\begin{equation}
\sigma(r)= \mbox{ } \sigma_{0} \mbox{ } \exp \mbox{ } [-(\frac{r}{r_{0}})^{n}],
\end{equation}

\noindent
in which $\sigma(r)$ is the surface brightness in linear units of luminous flux density at $r$,
$\sigma_0$ is the central surface brightness and $r_0$ is the angular scalelength.
The extrapolated central surface brightness is therefore:
$\mu_{0}=-2.5 \log_{10}\sigma_{0}$ in mag.arcsec$^{-2}$, whence
the equivalent expression in logarithmic surface-brightness units is: 

\begin{equation}
\mu(r)= \mbox{ } \mu_{0} \mbox{ } + \mbox{ } 1.086(\frac{r}{r_{0}})^{n},
\end{equation}

\noindent
enabling values of $\mu_0$ and $r_0$ to be obtained by linear regression when the optimum
value of $n$ has been derived. The analytical solution

\begin{equation}
2\pi\int_{0}^{\infty} \sigma_{0} r e^{-(\frac{r}{r0})^{n}} \mbox{ } dr=
\mbox{ } \frac{2}{n} \pi \sigma_{0} \Gamma(\frac{2}{n}) {r_{0}}^{2},
\end{equation}

\noindent
then yields an estimate of the total luminous flux within the pass-band concerned.

Clearly, this generalisation incorporates not only the $r^{\frac{1}{4}}$ law
($n=\frac{1}{4}$) but also exponentials ($n=1$) and Gaussians ($n=2$) as well
as both intermediate and more extreme cases.
Although it is a one-component model, galaxies exhibiting two-component
[or yet more complicated] structure can be comfortably accommodated by this scheme after 
their images have been smoothed sufficiently.

For the benefit of readers wishing to use this extrapolation system, a 
listing of the relevant {\scriptsize FORTRAN\/} code can be found in 
Appendix A. The subroutine {\scriptsize EXTRAPOL.FOR}:
(1) increments the profile-curvature parameter $n$, 
and attempts (for each $n$) to fit S\'{e}rsic's law to a surface-brightness profile defined by 
elliptical isophotes ( $r_i$, $\mu_i$($r$), $\sigma_{\mu_{i}(r)}$ ); 
(2) quantifies the quality of the fit obtained for each value of $n$; and
(3) integrates the volume beneath the surface defined by the best-fitting 
profile form (after rotation through 2$\pi$ radians about $r=0$) to $r=\infty$, 
thereby yielding a total-magnitude estimate.  
It calls the subroutine {\scriptsize FIT\/} and the function {\scriptsize
GAMMLN}, both from Press et al.\ (1986). Note that {\scriptsize FIT\/} 
actually fits a straight line of the form $Y=A+BX$, {\em not\/} one of the 
form $Y=AX+B$. Also, in order to reduce the dependence on other subroutines
and functions, we removed the lines:\\
{\scriptsize Q=1.} and
{\scriptsize Q=GAMMQ(0.5*(NDATA-2),0.5*CHI2)},\\
from this subroutine and removed the parameter 
{\scriptsize Q\/} from Line 1.

Although $t$-system total magnitudes cannot generally be derived directly from 
high resolution surface photometry of bright galaxies in the literature, they can be derived from
surface photometry of virtually all galaxies fainter than $B_{25}\sim15$ mag.\ provided that the 
resolution
of the photometry [after smoothing] is coarser than about 4.\uarcs5 (FWHM).
In the compilation of their VPC, Young \& Currie (in press) found that a reasonable amount
of smoothing of their plate-scan data was necessary simply in order to prevent the fragmentation of images during the
image segmentation process. They found that the minimum degree of smoothing required by the
segmentation software was actually sufficient for the derivation of $t$-system total magnitudes from
the surface-brightness profiles of all unsaturated galaxies with enough isophotes above the
limiting one for the fits to be performed.

\section{System stability}

In this section, we test the stability of the $t$ system with respect to
morphological type, size of atmospheric seeing disc, degree of smoothing and 
limiting isophote.
These tests are based on CCD surface photometry of four galaxies of different
morphological types. Details of the original observations and of the early 
stages of the reduction procedures are listed and/or referenced below for each 
galaxy individually.

{\em Spiral (Sbc) at:} $(\alpha,\delta)_{1950.0}=$22$^{\rm h}$01$^{\rm m}$06$^{\rm s}$, 
$-$20$^{\circ}$07\arcm24\arcsec. One 200-s $B$-band exposure of this galaxy was used; the 
frame having been taken in 1989 using the RCA CCD chip at the prime focus of 
the 2.5-m Isaac Newton Telescope. The FWHM of the seeing disc at the time of 
observation was 1.\uarcs1.
A 300$\times$300 0.\uarcs741-pixel subsection of the field was used after being
cleaned of stars and other galaxies. For details of the reduction and
calibration procedures adopted, see Metcalfe et al.\ (1995).

{\em Lenticular: NGC 7180.} One 600-s $B$-band exposure of this galaxy was used; the 
frame having been taken
in 1989 with the TI CCD chip on the 0.9-m telescope of the Cerro Tololo Inter-American
Observatory. The FWHM of the seeing disc was about 1.\uarcs5. The original 396$\times$396
array of 0.\uarcs494 pixels was binned up to a 197$\times$197 array of 0.\uarcs988
pixels [by omitting the peripheral pixels in the original array]. A 120$\times$120 
subsection of the resulting frame was then used after it had been cleaned of stars and 
other galaxies. For details of the reduction and calibration procedures adopted, 
see Metcalfe et al.\ (1995).

{\em Classical elliptical: NGC 6411.} One 200-s $B$-band exposure of this galaxy was used;
the frame having
been taken using a Loral CCD at the prime focus of the 4.2-m William Herschel Telescope
on La Palma. The observation was made at UT 22:08 on 1997 September 4, when the FWHM of
the seeing disc was 1.\uarcs1. Stellar images on the frame were removed using Starlink's
{\scriptsize GAIA\/} package and the photometric zero point was based on observations
of Landolt (1983) standard stars. The original 2048$\times$2048 array of 0.\uarcs26 pixels
was binned up to a 512$\times$512 array of 1.\uarcs02 pixels. A 360$\times$360 pixel
subsection of the binned-up array was then used. 

{\em Dwarf elliptical: NGC 147.} This galaxy is a member of the Local Group and lies 
within the vicinity of M31. Nine 600-s and four 900-s exposures of this galaxy were made 
of it using the CCD camera on the 60/90-cm F/3 Schmidt Telescope of Beijing Astronomical 
Observatory's (hereunder BAO) Xing Long Station. The observations were made between 
UT~14:03 and 17:08 on 1996 October 18, when the FWHM of the seeing disc was 2.\uarcs6. The 
CCD chip used was a Ford device which had an array size of 2048$\times$2048 pixels, and 
a corresponding field of view of 54\arcm 37\arcsec $\times$ 54\arcm 37\arcsec. In the 
absence of a broad-band filter, an $i$-band Beijing-Arizona-Taipei-Connecticut 
(hereunder BATC) survey filter was employed. The BATC filter system is
described at length by Fan (1995) and briefly by Fan et al.\ (1996) who
refer to the $i$-band filter as Filter No.~9. This filter's transmission
curve peaks at a wavelength of 6600\AA and has a FWHM of 480\AA.
After bias subtraction, flat fielding and the removal of spurious images
caused by cosmic-ray events; all thirteen CCD frames were stacked, thereby yielding
a single frame whose effective integration time was 9000 s. The procedure adopted
for these reductions was the same as adopted by Fan et al.\ (1996).
As Hodge (1976) found no evidence for any global colour 
gradient in NGC~147, we were able to transform the $i$-band images directly to 
the $B$ system by calibration with the $r_{\rm TG}$-band surface-brightness 
profile of Kent (1987) and
the transformation, $B=r_{\rm TG}+1.21$ from Young \& Currie (1994).
After the calibration process, the stacked frame was binned up to one with
15.\uarcs03 pixels. 

Although NGC~6411 and NGC~147 were not observed by
Metcalfe et al.\ (1995), the same software as used by those authors was applied to the
reduced but unsmoothed frames of these galaxies in order to generate Kron-system total magnitudes 
for them. The values obtained were $B_K$=12.88 and 10.36 respectively.

\begin{figure*}
\psfig{figure=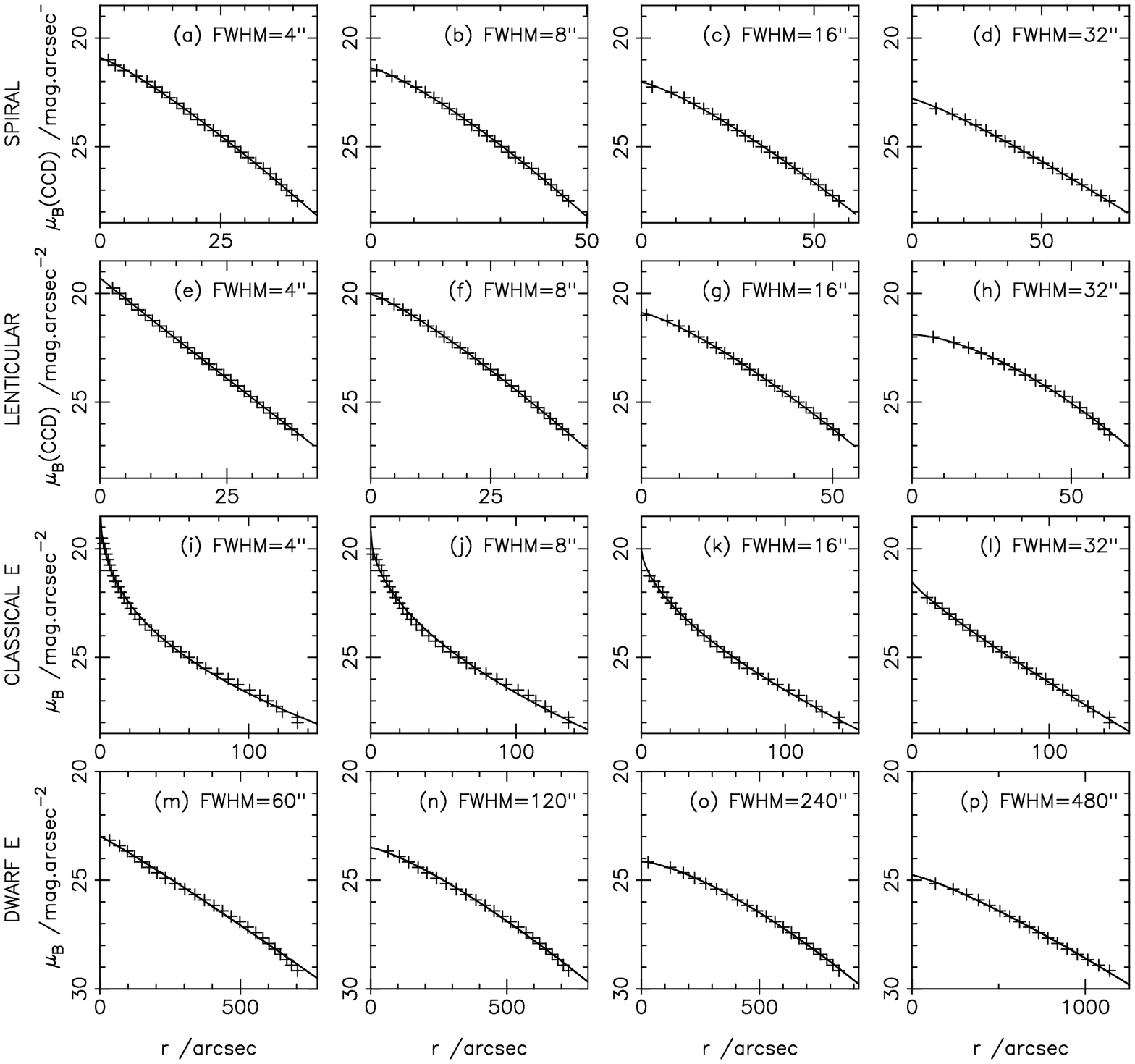,height=17cm,angle=-90}
\caption{The simulated effect of poor to very poor seeing conditions on the 
surface-brightness profiles of bright galaxy images:
{\bf (a, b, c and d)\/} a spiral, {\bf (e, f, g and h)\/} a lenticular, 
{\bf (i, j, k and l)\/} a classical elliptical and 
{\bf (m, n, o and p)\/} a dwarf-elliptical galaxy. This effect is analogous to
the effect of ordinary seeing conditions on more distant galaxies of the
same type and physical size.  
The image resolution function adopted was that of Moffat (1969) and the
FWHM of the synthetic seeing discs are shown in arcsec. 
The curves represent model S\'{e}rsic profiles fitted to all plotted 
isophotes.}
\label{profs}
\end{figure*}

\begin{table*}
\caption{Best-fitting S\'{e}rsic model parameters for the synthetic seeing-distorted 
surface-brightness profiles depicted in Fig.~1, and for the same profiles
but with brighter limiting isophotes}
\begin{center}
\begin{tabular}{rrrrrrrrr}
\hline
\noalign{\smallskip}
type & seeing      & limiting isophote  & \multicolumn{3}{c}{model parameters} & $m_t$ & \multicolumn{2}{c}{quality of fit$^{\rm a}$}\\
     & FWHM/arcsec & /mag.arcsec$^{-2}$ &   $n$   &  $\mu_{0}$  &   $r_0$   & /mag.\ & $\chi^{2}$ & $\nu$ \\
\noalign{\smallskip}
\hline
\noalign{\smallskip}
S    &  4          & $\mu_{B}$(CCD)=25.50 & 1.20    & 20.90       & 0.920E+01 & 14.40  & 3.5961 & 17 \\
     &  8          & \arcsec              & 1.35    & 21.43       & 0.125E+02 & 14.41  & 0.8579 & 15 \\
     & 16          & \arcsec              & 1.39    & 22.11       & 0.172E+02 & 14.42  & 1.0134 & 12 \\
     & 32          & \arcsec              & 1.41    & 22.97       & 0.256E+02 & 14.44  & 0.1360 &  8 \\
S    &  4          & 26.50                & 1.18    & 20.89       & 0.905E+01 & 14.40  & 3.9638 & 21 \\
     &  8          & \arcsec              & 1.29    & 21.39       & 0.120E+02 & 14.40  & 2.1809 & 19 \\
     & 16          & \arcsec              & 1.30    & 22.06       & 0.164E+02 & 14.41  & 2.2567 & 16 \\
     & 32          & \arcsec              & 1.26    & 22.87       & 0.235E+02 & 14.40  & 0.5753 & 12 \\
S    &  4          & 27.50                & 1.20    & 20.91       & 0.922E+01 & 14.40  & 4.6817 & 25 \\
     &  8          & \arcsec              & 1.29    & 21.40       & 0.120E+02 & 14.40  & 2.2766 & 23 \\
     & 16          & \arcsec              & 1.27    & 22.03       & 0.160E+02 & 14.40  & 2.6476 & 20 \\
     & 32          & \arcsec              & 1.18    & 22.80       & 0.219E+02 & 14.38$^{\rm b}$  & 1.0317 & 16 \\
\\
S0   &  4          & $\mu_{B}$(CCD)=24.50 & 0.93    & 19.20       & 0.521E+01 & 13.46  & 0.6036 & 18 \\
     &  8          & \arcsec              & 1.16    & 19.99       & 0.907E+01 & 13.47  & 0.5109 & 16 \\
     & 16          & \arcsec              & 1.34    & 20.92       & 0.150E+02 & 13.49  & 1.5291 & 13 \\
     & 32          & \arcsec              & 1.45    & 21.82       & 0.240E+02 & 13.45  & 0.1496 &  9 \\
S0   &  4          & 25.50                & 0.96    & 19.25       & 0.549E+01 & 13.47  & 1.7091 & 22 \\
     &  8          & \arcsec              & 1.19    & 20.02       & 0.934E+01 & 13.47  & 0.9366 & 20 \\
     & 16          & \arcsec              & 1.29    & 20.89       & 0.146E+02 & 13.48  & 2.3516 & 17 \\
     & 32          & \arcsec              & 1.48    & 21.83       & 0.243E+02 & 13.46  & 0.3424 & 13 \\
S0   &  4          & 26.50                & 0.98    & 19.28       & 0.569E+01 & 13.47  & 2.2644 & 26 \\
     &  8          & \arcsec              & 1.21    & 20.04       & 0.952E+01 & 13.47  & 1.5676 & 24 \\
     & 16          & \arcsec              & 1.29    & 20.89       & 0.146E+02 & 13.48  & 2.5165 & 21 \\
     & 32          & \arcsec              & 1.60    & 21.90       & 0.257E+02 & 13.47$^{\rm b}$  & 2.7139 & 17 \\
\\
E    &  4          & $\mu_{B}$=26.00      & 0.39    & 17.72       & 0.443E+00 & 12.81  & 45.6758 & 25 \\
     &  8          & \arcsec              & 0.55    & 19.33       & 0.296E+01 & 12.86  & 44.8807 & 22 \\
     & 16          & \arcsec              & 0.64    & 20.15       & 0.613E+01 & 12.86  &  4.9034 & 18 \\
     & 32          & \arcsec              & 0.85    & 21.50       & 0.181E+02 & 12.84  &  0.7091 & 14 \\
E    &  4          & 27.00                & 0.37    & 17.54       & 0.323E+00 & 12.78  & 48.0880 & 29 \\
     &  8          & \arcsec              & 0.51    & 19.17       & 0.228E+01 & 12.81  & 53.6671 & 26 \\
     & 16          & \arcsec              & 0.60    & 20.01       & 0.507E+01 & 12.82  &  7.3776 & 22 \\
     & 32          & \arcsec              & 0.84    & 21.48       & 0.176E+02 & 12.84  &  0.7639 & 18 \\
E    &  4          & 28.00                & 0.38    & 17.63       & 0.380E+00 & 12.80            & 51.2237 & 33 \\
     &  8          & \arcsec              & 0.51    & 19.17       & 0.229E+01 & 12.81            & 54.7888 & 30 \\
     & 16          & \arcsec              & 0.60    & 20.01       & 0.507E+01 & 12.82            &  8.2497 & 26 \\
     & 32          & \arcsec              & 0.86    & 21.52       & 0.185E+02 & 12.84$^{\rm b}$  &  1.7321 & 22 \\
\\
dE   &  60         & $\mu_{B}=27.16$      & 0.89    & 22.73       & 0.110E+03 & 10.27  & 1.1096 & 15 \\
     & 120         & \arcsec              & 1.07    & 23.27       & 0.166E+03 & 10.29  & 0.2913 & 13 \\
     & 240         & \arcsec              & 1.35    & 24.09       & 0.278E+03 & 10.33  & 0.1603 & 11 \\
     & 480         & \arcsec              & 1.47    & 24.94       & 0.415E+03 & 10.40  & 0.0156 &  7 \\
dE   &  60         & 28.16                & 0.96    & 22.83       & 0.123E+03 & 10.30  & 3.0475 & 19 \\
     & 120         & \arcsec              & 1.17    & 23.37       & 0.185E+03 & 10.32  & 1.3257 & 17 \\
     & 240         & \arcsec              & 1.38    & 24.10       & 0.283E+03 & 10.33  & 0.2436 & 15 \\
     & 480         & \arcsec              & 1.37    & 24.89       & 0.397E+03 & 10.37  & 0.1210 & 11 \\
dE   &  60         & 29.16                & 1.07    & 22.98       & 0.145E+03 & 10.31  & 9.2790 & 23 \\
     & 120         & \arcsec              & 1.29    & 23.50       & 0.207E+03 & 10.32  & 4.0134 & 21 \\
     & 240         & \arcsec              & 1.43    & 24.13       & 0.290E+03 & 10.34  & 0.8021 & 19 \\
     & 480         & \arcsec              & 1.22    & 24.77       & 0.357E+03 & 10.34$^{\rm b}$  & 0.8673 & 15 \\
\noalign{\smallskip}
\hline
\end{tabular}
\end{center}
\begin{list}{}{}
\item$^{\rm a}$ $\nu$ represents degrees of freedom (number of isophotes minus two)
\item$^{\rm b}$ These values were (after system transformation when relevant) adopted in Table~2
\end{list}
\label{fits}
\end{table*}

For each galaxy, four synthetic low-resolution images were generated. This involved the
convolution of each original or stacked image with Moffat (1969) functions of 
($\sqrt{16-d^2}$)\arcsec, 8\arcsec, 16\arcsec and 32\arcsec FWHM
in the case of the classical galaxies or with the same functions
of 1\arcm, 2\arcm, 4\arcm and 8\arcm FWHM, in the case of
NGC~147; where $d$ was the FWHM of seeing disc at 
the time of the original observation(s). In order to minimize edge effects, each
digital image was embedded in a very much larger array of pixels (in which each 
pixel in the surrounding grid was set to zero)
before any convolution was performed.

The Moffat function was chosen on this occasion in order to simulate both the effect 
of poor seeing on nearby objects and the effect of average
seeing on distant objects. Note that in adopting a Moffat function here, 
we are actually applying a very much more stringent test on the stability of 
the $t$ system than we would have been had we adopted the Gaussian function
that we recommend for the purpose of smoothing. This is because the Moffat 
function is a much more complicated function than the Gaussian one, being
similar to the Gaussian at small radial distances but falling off
much more slowly at larger radial distances.

Godwin's (1976) image-segmentation software, as outlined by Carter \& Godwin
(1979), was used in order to fit elliptical isophotes of 0.25 mag.arcsec$^{-2}$
separation (each defined by a mean radius $r$, an ellipticity and a position 
angle) to all of the synthetic low-resolution images. The isophotes were
weighted according to the simple algorithm: $\sigma_{\mu}=0.05$ for 
$\mu\leq20.0$ or $\sigma_{\mu}=0.02(\mu-20.0)$ for $\mu>20.0$.
The resulting synthetic surface-brightness profiles are plotted in 
Fig.~\ref{profs} together with the best fitting S\'{e}rsic model profiles; 
whilst the corresponding model parameters are tabulated in Table~\ref{fits}, which
also lists those model parameters obtained when different limiting isophotes
were applied.

As can be seen from Fig.~\ref{profs} and Table~\ref{fits}, S\'{e}rsic's model
yields very consistent results for all of the synthetic images except for the 
two highest resolution images of the classical elliptical (which, at
12th magnitude is in fact a very bright object)
and the highest resolution image of the dwarf elliptical. In these three cases 
one-component profile models appear not to be completely adequate. 
However, once the resolution of a 
galaxy image has been degraded sufficiently, even if purely by seeing effects,
the $t$-system total magnitudes obtained do appear to be stable
typically to a couple of percent or so, irrespective of morphological type, 
the size of the seeing disc, or the limiting isophote--provided that the limiting 
isophote is not so bright that there are too few isophotes to fit.

Note that under normal circumstances, when the seeing disc is {\em not\/} 
almost as large as the galaxy image itself and the resolution of the image can 
be deliberately degraded by convolution with a Gaussian function (or even
a simple Hanning function), the level of stability with respect to image
resolution must be even greater than this. This is because the synthetic 
surface-brightness profile obtained by convolving any galaxy image 
with a Gaussian function of large FWHM, must be more Gaussian than the 
original profile and therefore more likely to be well described by 
S\'{e}rsic's law (as the Gaussian function, unlike the Moffat 
function for example, can be perfectly described by S\'{e}rsic's law).

We also tested the system for stability with respect to different
weighting schemes for the isophotes, and found that whilst altering the 
weightings had very significant effects on the $\chi^{2}$ values obtained,
and reasonably significant effects on which best-fitting parameters were
adopted, the effects on the total magnitude values obtained were only
at the one or two per cent level--for realistic weighting schemes at least.

\section{Comparisons with other systems}

\subsection{High-resolution images}

\begin{table}
\caption{A comparison between Johnson $B$-band total-magnitude values obtained from high-resolution
galaxy images using different extrapolation systems}
\begin{center}
\begin{tabular}{llrrrrrr}
\hline
\noalign{\smallskip}
type & designation      & $(B_T)^a$ & $(m_B)^b$ & $(B_K)^c$ & $(B_t)^d$ \\
     & or $(\alpha,\delta)_{2000.0}$&           &           &           \\ 
\noalign{\smallskip}
\hline
\noalign{\smallskip}
Sbc  & (22:03:51.2,     &   N/A     & 14.41     & 14.49     & 14.47 \\
     & $-$19:52:51)     &           &           &           &       \\
S0   & NGC~7180         & 13.56     & 13.61     & 13.67     & 13.65 \\
E    & NGC~6411         & 12.79     & 12.93     & 12.88     & 12.84 \\
dE   & NGC~147          & 10.47     & 10.43     & 10.36     & 10.34 \\
\noalign{\smallskip}
\hline
\end{tabular}
\end{center}
\begin{list}{}{}
\item$^{\rm a}$ $T$ system extrapolation of aperture and/or surface photometry, RC3
\item$^{\rm b}$ Zwicky magnitude transformed to $T$ system, RC3
\item$^{\rm c}$ Kron system extrapolation, Metcalfe et al.\ (1995) or this work
\item$^{\rm d}$ $t$ system extrapolation, this work
\end{list}
\label{comp}
\end{table}

In Table~\ref{comp}, we have transformed those CCD-system total-magnitude 
estimates flagged with a superscript `b' symbol in Table~\ref{fits} into Johnson or Cousins
system magnitudes based on the colour equations of Metcalfe at al.\ (1995).
Note that we did {\em not\/} invoke Metcalfe et al.'s magnitude values for 
any of the relevant galaxies, {\em only\/} their colour values when necessary.

It is clear from Table~\ref{comp} that the agreement between the
different systems for the objects considered is excellent. Note that for the
three classical galaxies, the
zero points on which the Kron-system and $t$-system values were based were
the same, whilst those on which the $T$-system values were based were 
independent. In the case of the dwarf elliptical though, the $T$-system and
$t$-system values were both based on the zero point of Kent (1987).

For the sake of completeness, we would very much have liked to include
a dwarf galaxy whose $n$ value is much greater than 1.0 in the comparisons
performed in this subsection. However, we have not yet been able to obtain
deep CCD images of a suitable galaxy. In any case, as mentioned in Section~2,
there can be no doubt that such objects cannot be accommodated by the $T$ system,
which necessarily over-estimates their luminosities.

\subsection{Low-resolution images}

Whilst the agreement between the three systems is very good for the four resolved 
galaxy images already investigated, let us now consider what would happen 
if we attempted to estimate the total magnitudes of these same galaxies if 
they were, hypothetically, re-located at much greater distances from our 
galaxy. Clearly, seeing effects would become more significant than they
were during the original observations. In fact, the effects can be understood
from Fig~\ref{profs}, if one interprets greater FWHM values as the same
degree of seeing due to a particular galaxy being re-located to greater
distances from us, and one re-scales the absolute radial-distance scales accordingly.
For example, in the case of NGC~147, which is about 0.67 Mpc
distant (Lee et al.\ 1993), Fig.~1(i-l inclusive) represents profiles under purely
hypothetical atmospheric conditions in which the FWHM of the seeing discs
are 60-480\arcsec, but also represents the same galaxy if re-located at a
distance of 13.6 Mpc and observed under conditions with seeing discs of
0.\uarcs3-2.\uarcs4 FWHM. As one would expect, the profiles of more distant and/or 
physically smaller galaxies are more susceptible to distortion by seeing
effects than those of nearby and/or physically larger systems. 

\begin{table}
\caption{$T$-system total magnitude estimates based on the synthetic classical elliptical 
galaxy profiles$^{\rm a}$ plotted in Fig.~\ref{profs} as a function of the FWHM of the seeing disc}
\begin{center}
\begin{tabular}{rrrrr}
\hline
\noalign{\smallskip}
limiting isophote  & \multicolumn{4}{c}{FWHM of seeing disc /arcsec} \\ 
$\mu_B$/mag.arcsec$^{-2}$         &     4 &     8 &     16 &     32 \\
\noalign{\smallskip}
\hline
\noalign{\smallskip}
26.0 & 12.57 & 12.39 & 12.45 & 12.27 \\
27.0 & 12.63 & 12.52 & 12.57 & 12.45 \\
28.0 & 12.71 & 12.65 & 12.65 & 12.55 \\
\noalign{\smallskip}
\hline
\end{tabular}
\end{center}
\begin{list}{}{}
\item$^{\rm a}$ For the sake of consistency, we used the same isophotal weighting scheme
as for the $t$-system profile fitting procedure 
\end{list}
\label{r^1/4}
\end{table}

In the case of the $t$ system, distortion of image profiles due to seeing effects 
can be accounted for, as was demonstrated in Section~4. However, because
the $T$ system assumes that a galaxy's surface-brightness profile [or the 
integrated luminosity equivalent] is only a function of morphological type and not
of image resolution, it is therefore only applicable to highly-resolved
galaxy images. As is evident from Table~\ref{r^1/4}, if seeing effects are
significant but not taken into account, this will generally result in an
$over$-estimate of luminosity for a particular galaxy. 

The Kron system, by contrast, does not make any prior assumption as to the
profile shape of a target galaxy, though it does make a smaller assumption as to
the shape of the curve representing the logarithmic derivative of the 
light-growth at large radial distances. We would therefore expect Kron-system
total-magnitude scales to be very stable with respect to the size of the seeing disc, but
there may of course still be room for small second order effects due to the assumption
mentioned.

\subsection{Images due to point sources}

\begin{table}
\caption{Estimates of the total light contained under two typical [two-dimensional radially symmetric] 
Moffat surfaces, obtained by fitting different profile laws to different numbers of isophotal levels; 
the resulting values being quoted as linear fractions of the actual total-light values (which have both
been set to unity)
}
\begin{center}
\begin{tabular}{lrrr}
\hline
\noalign{\smallskip}
law adopted & no.\ isophotal    & Hamburg & Lick     \\ 
            & levels$^{\rm a}$ & Schmidt & 120-inch \\
\noalign{\smallskip}
\hline
\noalign{\smallskip}
de Vaucouleurs $r^{\frac{1}{4}}$  &  8 & 6.391 & 6.315 \\  
\arcsec                           & 16 & 1.823 & 1.808 \\
\arcsec                           & 24 & 1.443 & 1.363 \\
\arcsec                           & 32 & 1.552 & 1.349 \\
\\
S\'{e}rsic $r^n$                  &  8 & 0.933 & 0.881 \\
\arcsec                           & 16 & 0.978 & 0.955 \\
\arcsec                           & 24 & 0.995 & 0.988 \\
\arcsec                           & 32 & 1.001 & 1.003 \\
\\
Gauss $r^2$                       &  8 & 0.885 & 0.812 \\
\arcsec                           & 16 & 0.963 & 0.938 \\
\arcsec                           & 24 & 1.002 & 1.033 \\
\arcsec                           & 32 & 1.015 & 1.120 \\
\noalign{\smallskip}
\hline
\end{tabular}
\end{center}
\begin{list}{}{}
\item$^{\rm a}$ The depth of the simulated surface-photometry in terms of the number of isophotal levels 
(of 0.25~mag.arcsec$^{-2}$ separation) fitted
\end{list}
\label{fitmoffat}
\end{table}

In order to test the applicability of the $t$ system to images due to point sources, we simulated the
surface photometry one might expect to obtain for two model point spread functions. The two model profiles
adopted were based on Moffat functions using the original parameter values quoted by Moffat (1969) for traced
stellar images on Hamburg Schmidt and the Lick 120-inch reflector photographic plates. The adopted parameter
values were $\beta=4$ and $R=5\times10^{-5}$m for the Hamburg plates and $\beta=2.72$, and $R=1.12\times10^{-4}$m
for the Lick plates. The brightest isophote in each case was taken to correspond to that radial distance, $r$,
this time in metres,
at which the surface brightness was 0.125 mag.m$^{-2}$ fainter than the peak value at $r=0$. Radii were then
computed for further [circular] isophotes that were multiples of 0.25 mag.m$^{-2}$ fainter than the brightest 
isophote.

The results of attempts to fit not only S\'{e}rsic's law, but also the Gaussian and $r^{\frac{1}{4}}$ laws,
to the brighter isophotes of these model profiles are shown in Table~\ref{fitmoffat}. 
For these fits, the adopted weighting scheme was: $\sigma_{\mu}=0.05+0.02(\mu_{0}-\mu)$, where $\mu_0$ is the
peak central surface brightness at $r=0$ in units of mag.m$^{-2}$. Note that although Moffat did 
not quote plate-scale values, we were still able to investigate the {\em fractional\/} differences between the 
extrapolated total-luminosity estimates and the true luminosities represented by the two-dimensional
radially-symmetric Moffat surfaces\footnote{The latter surfaces having been integrated to very large radial 
distances using Simpson's rule with very small radial distance intervals.}.

From Table~\ref{fitmoffat}, it is clear that, as one would expect, the $r^{\frac{1}{4}}$ law always severely
over-estimates the light due to a point-spread function represented by a Moffat profile. It is also evident from 
the same table that an extrapolation of S\'{e}rsic's model generally yields a significantly better estimate of 
the total light due to a typical point-spread function than does an extrapolation of a strictly Gaussian model. 

In the above comparisons, we have of course applied a very stringent test to $t$-system and Gaussian-law
extrapolations, by invoking pure Moffat functions to describe the unresolved images. In practice, as in
the compilation of the VPC for example, the structures of the unresolved images should really be
described by the product of three functions, the sampling function 
a Moffat function and the smoothing function, all convolved with one another.
Provided that the smoothing function is a single-component function that falls off steeply with
increasing radial distance, such as a Gaussian function, better results should be obtainable 
using S\'{e}rsic's model (or even a strictly Gaussian model) than
those tabulated in Table~\ref{fitmoffat}. This is because a pure Moffat function is less amenable to
being described by a S\'{e}rsic function (or the Gaussian case thereof) than is a Moffat function that has been
deliberately smoothed. The $t$ system is therefore applicable to point-source galaxy images as well as 
resolved-galaxy images. While it cannot offer a perfect fit to a typical unsmoothed 
point-spread function, it does generally offer significantly better results
than those that can be obtained by invoking a 
purely Gaussian model. 

\section{Summary}

We have presented a new procedure for obtaining total-magnitude 
estimates from unsaturated galaxy images. This method involves first 
smoothing two-dimensional digital images numerically (ideally convolving them
with a two-dimensional radially symmetric Gaussian function of sufficient 
FWHM), in order to produce lower-resolution images, which are then
parameterised using elliptical isophotes. The rationale behind this
is that even the faintest isophotes of the synthetic lower-resolution images (which
cannot be measured accurately or at all in practice) become distorted in a 
predictable manner by the smoothing process. The second stage involves modeling the
resulting surface-brightness profiles with S\'{e}rsic functions, and 
extrapolating the best-fitting functions to infinite radial distances.
We have also demonstrated the system's high level of stability with respect to
galaxy morphological type, limiting isophote and the size of the seeing disc (and 
thereby degree of smoothing too [provided one does not under-smooth],
as typical smoothing functions cause distortions that
are more easily accommodated by the fitting procedure than those distortions caused purely by
seeing effects). 

\begin{acknowledgements}
We would like to thank Jon Godwin for providing the photometry software
and Steve Maddox for a useful discussion. 
This work made use of the computing facilities of
the QSO \& Observational Cosmology Group at BAO and those of Starlink.
CKY gratefully acknowledges a PDRF from the National Postdoctoral 
Fellowship Office of China.

\end{acknowledgements}

\vspace{10mm}

\noindent
{\bf Appendix A: FORTRAN code}

\begin{verbatim}
      SUBROUTINE EXTRAPOL
     : (RMEANR,YMU,SIG,NDATA,
     : SN,SMU0,SR0,TOTAL,SCHI2,NU)
* -------------------------------------------
* This subroutine fits Sersic profile parame-
* ters to a surface-brightness profile by chi
* squared minimisation. It then extrapolates 
* the profile to obtain a t-system total mag-
* nitude estimate. 
* -------------------------------------------
* INPUT PARAMETERS (all unchanged on output):
*   RMEANR(100): mean radial distance of each 
* isophote /arcsec;
*   YMU(100): corresponding surface bright- 
* ness of each isophote /mag.arcsec**(-2);
*   SIG(100): corresponding 1 sigma uncer-
* tainty on each YMU;     
*   NDATA: number of isophotes (<101);
* -------------------------------------------
* OUTPUT PARAMETERS:
*   SN, SMU0 and SR0: best fitting Sersic
* parameters: n, mu_0 and r_0 respectively
* (see Equation 6);
*   TOTAL: t-system total magnitude derived
* from the best fitting parameters;
*   CHI2: chi squared value for adopted fit;
*   NU: corresponding degrees of freedom.
* -------------------------------------------
* Two external routines called from Numerical 
* Recipes, Press et al., Cambridge U.P. 1986:
*   SUBROUTINE FIT (with the minor modifica-
* tions described in Section 3) and
*   FUNCTION GAMMLN (without modification).
* -------------------------------------------
      REAL RMEANR(100), YMU(100), SIG(100),
     : XMEANRN(100)
      INTEGER NDATA(100)
      DOUBLE PRECISION DR0, DNP1, GAMMLN
      PI= 3.141592654
      BCHI2= 100000.0
      MWT= 1
* increment n from 0.2 to 3.0
        DO I= 20,300
          RN= 0.01*FLOAT(I)
            DO J= 1,NDATA
              XMEANRN(J)= (RMEANR(J))**RN
            ENDDO
          CALL FIT (XMEANRN,YMU,NDATA,SIG,MWT,
     :     FMU0,FSLOPE,SIGMU0,SIGS,CHI2)
* retain parameters of best fit so far
            IF (CHI2.LE.BCHI2) THEN
              BESTN=  RN
              BMU0=   FMU0
              BSLOPE= FSLOPE
              BCHI2=  CHI2
            ENDIF
        ENDDO
      SCHI2= BCHI2
      NU= NDATA-2
      SN= BESTN
      SMU0= BMU0
      SR0= (1.086/BSLOPE)**(1.0/SN)
* evaluate total magnitude      
      DR0= DBLE(SR0)
      DNP1= (2.0D0/DBLE(SN))
      GAMP1= EXP(SNGL(GAMMLN(DNP1)))
      TOTAL= -2.5*LOG10(2.0*PI*GAMP1/SN)
     : +SMU0-2.5*SNGL(DLOG10(DR0**2D0))
      END
* -------------------------------------------
\end{verbatim}

\end{document}